\documentclass[letterpaper,twoside]{JHEP3}

\usepackage{epsfig}
\newcommand{\fa}{f_{a}}
\newcommand{\g}{g_{a \gamma}}
\newcommand\f{f_{12}}
\newcommand{\den}{{\rm g}\,{\rm cm}^{-3} }

\newcommand{\s}{{\rm s}}
\newcommand{\R}{{d_{\rm SN}}}

\newcommand\eV{\mbox{eV}}
\newcommand\keV{\mbox{keV}}
\newcommand\MeV{\mbox{MeV}}
\newcommand\GeV{\mbox{GeV}}
\title{New constraints for heavy axion-like particles from supernovae}
\author{M. Giannotti$^a$, L.D. Duffy$^b$ and R. Nita$^a$\\\llap{$^a$}Physical Sciences, Barry University,\\ Miami Shores, FL 33161, U.S.A,\\\llap{$^b$}Los Alamos National Laboratory,\\Los Alamos, NM 87545, U.S.A.\\E-mail: \email{mgiannotti@mail.barry.edu, ldd@lanl.gov, Rafaela.Nita@mymail.barry.edu}}

\abstract{We derive new constraints on the coupling of heavy pseudoscalar (axion-like) particles to photons, based on the gamma ray flux expected from the decay of these particles into photons. After being produced in the supernova core, these heavy axion-like particles would escape and a fraction of them would decay into photons before reaching the Earth. We have calculated the expected flux on Earth of these photons from the supernovae SN 1987A and Cassiopeia A and compared our results to data from the Fermi Large Area Telescope. This analysis provides strong constraints on the parameter space for axion-like particles.  For a particle mass of $100\,$MeV, we find that the Peccei-Quinn constant, $\fa$, must be greater than about $10^{15}\,\GeV$.  Alternatively, for $\fa=10^{12}\,\GeV$, we exclude the mass region between approximately $100\,\eV$ and $1 \,\GeV$.}

\keywords{axions, axion-like particles, supernovae}

\preprint{LA-UR 10-05895}

\begin{document}

\section{Introduction}

In this work, we analyze the gamma ray flux expected on Earth from the decay of heavy pseudoscalar particles produced in supernovae (SNe). 
The analysis of this flux and comparison with existing experimental observations
provide new limits on these particles for the mass range from $\sim 100\,\eV$ to $\sim 1\,$GeV.    

Pseudoscalar particles occur in many extensions of the standard model. 
Arguably, the best known pseudoscalar is the 
axion~\cite{Peccei:1977hh,Peccei:1977ur,Weinberg:1977ma,Wilczek:1977pj}, 
whose existence is a natural consequence of the Peccei-Quinn (PQ) 
mechanism for solving the strong CP problem. 
The axion emerges as a pseudo-Nambu-Goldstone boson from the axial Peccei-Quinn (PQ) symmetry, $U_{PQ}(1)$,
spontaneously broken at the PQ (or axion) scale, $\fa$. 
This scale determines the strength of the axion interactions, as can be seen from the Lagrangian terms,
\begin{equation}  \label{a-couplings}
\mathcal{L}_{\rm int} = -ig_{ak}a \bar\psi_k\gamma_5\psi_k +
{\frac{g_{a\gamma }}{4}} aF_{\mu\nu}\tilde F^{\mu\nu} + .... \, .
\end{equation}
In Eq.~(\ref{a-couplings}), $g_{ak}=C_km_k/\fa$ is the axion-fermion coupling, with $m_k$ the fermion mass,
and $\g=C_\gamma \alpha/(2\pi\fa)$ is the axion-photon coupling, where $\alpha$ is the fine structure constant. 
The constants $C_i$ are model-dependent parameters typically, but not necessarily, of order 1.

In the original axion models~\cite{Weinberg:1977ma,Wilczek:1977pj,Kim:1979if,Shifman:1979if,Zhitnitsky:1980tq,Dine:1981rt}, the axion mass is related to the PQ constant via
\begin{equation}\label{mass}
	m_a\simeq 6.0\;\mu\mathrm{eV}\left(\frac{f_a}{10^{12}\;\mathrm{GeV}}\right)^{-1}\,.
\end{equation}
These \textit{standard axion} models have been studied extensively in the past 30 years 
(see, e.g., ref.~\cite{Kim:2008hd} for a recent review). 
A pseudoscalar particle which possesses some, but not all, of the properties of the standard axion 
%interact through the Lagrangian terms of Eq.~(\ref{a-couplings}), but for which 
%relation (\ref{mass}) is not satisfied, 
is known as an \textit{axion-like} particle.  These particles have recently been the subject of considerable interest
(see, e.g.,~ref.~\cite{Jaeckel:2010ni} for a recent review). 

For the standard axion, experimental and astrophysical constraints exclude values of the PQ constant 
below $10^9\, \GeV$ \cite{Raffelt:1996wa}. 
By Eq.~(\ref{mass}), the axion is then very light, with $m_a\lesssim 10^{-2}$ eV
(see refs.~\cite{Kim:2008hd,
%Kim:1986ax,Cheng:1987gp,Raffelt:1996wa,Sikivie:2006ni,
Duffy:2009ig} for recent discussions). 
However, the existence of a heavy pseudoscalar particle is not excluded if relation (\ref{mass}) is relaxed. 
Several models for these \textit{heavy-axion-like} particles (HALPs) have been proposed in the past.
Examples include models of large extra  dimensions~\cite{ArkaniHamed:1998rs,ArkaniHamed:1998nn,Dienes:1999gw,DiLella:2000dn}, 
where heavy HALPs emerge as Kaluza-Klein (KK) modes of the standard \textit{light} axion, models
with Plank-scale induced mass terms \cite{Berezhiani:1999qh}
and models of axions interacting with a hidden sector~\cite{Rubakov:1997vp,Berezhiani:2000gh,Gianfagna:2004je}.
It is, therefore, phenomenologically interesting to explore the axion parameter space relaxing relation (\ref{mass}).

Observe that in some cases, for example in the models~\cite{Berezhiani:2000gh,Gianfagna:2004je}, the HALP is a QCD axion, 
which effectively solves the strong CP-problem. Therefore, the term \textit{axion-like} particle could be somewhat misleading. 
However, in general, these heavy pseudoscalar particles are not responsible for the solution of the strong CP problem,
and cannot be considered QCD axions.
%
%This is not always the case. 
%In the model of large extra  dimensions, \cite{ArkaniHamed:1998rs,ArkaniHamed:1998nn,Dienes:1999gw,DiLella:2000dn}, 
%the HALPs are not directly responsible for the solution of the strong CP-problem, 
%although they are related to the standard QCD axion, whereas in other models, such as \cite{Berezhiani:1999qh},
%the HALPs have different origins and are not related to the strong CP problem. 
%In any case, it is phenomenologically interesting to explore the axion parameter space relaxing relation (\ref{mass}). 

%The current exclusion limits on the \textit{axion-like} particle parameter space, 
%extending it beyond the standard axion window, can be found in ref.~\cite{Jaeckel:2010ni}.

Rather than focusing on the specific models, in this work we keep a phenomenological approach, 
and consider a general HALP, described by two independent parameters, the mass, $m_a$, and the PQ scale, $\fa$. 
We assume that this particle interacts with nucleons and photons as described by Eq.~(\ref{a-couplings})
with $C_{N}=1$ and $C_\gamma=0.7$. 
Our normalization choice for $C_\gamma$ is that of a DFSZ axion~\cite{Zhitnitsky:1980tq,Dine:1981rt} 
and is used in order to enable comparison with the existing literature.  
For the sake of simplicity, we neglect a possible HALP-lepton interaction.  As we discuss later, 
the existence of such an interaction would not significantly change our results. 

Presently, the strongest constraints on such HALPs %($m_a\geq 1\keV$) 
are derived from the analysis of the effects of the photons from HALPs decay
on some cosmological observations such as the extragalactic photon background, 
the cosmic microwave background, and the deuterium abundance~\cite{Masso:1997ru}. 
This argument is based on the hypothesis that the HALPs are 
only allowed to decay into photons.
%In general, however, they also have to couple to other matter, for example to leptons or nucleons, 
%in order to be produced in the early universe. 
The analysis applies to the mass region between 1~keV and 1~GeV, 
and $\fa$ is restricted to being greater than $10^{10}\GeV$ (see Fig.~1 in ref.~\cite{Masso:1997ru}) 
for an axion-photon coupling with $C_\gamma=0.7$. 

In this paper, we extend the bound on $\fa$ to greater values
for axion masses in a similar range, using Fermi-LAT observations of supernovae photons. 

Weakly interacting pseudoscalars may be produced via 
nucleon bremsstrahlung 
%\cite{Iwamoto:1984ir, Turner:1988bt, Brinkmann:1988vi, Raffelt:1993ix,Hanhart:2000ae,Giannotti:2005tn} 
in the core of supernovae.  
Due to their weak interactions, they would freely stream out and eventually decay into photons.  
The decay probability depends on both the HALP's mass and coupling with photons. 
In general, a fraction of the HALPs produced would decay before reaching Earth, generating an observable gamma-ray flux.  
The signal, peaked at energies of the order of the SN core temperature, 
would be stretched on a much longer timescale than the corresponding neutrino signal, 
and so the decay photons would  still be observable today.
%
%by several years, 
%relative to the light or neutrino signal produced directly in the supernova event. 

We have analyzed the gamma-ray signal expected from the decay of massive HALPs produced by SN 1987A and Cassiopeia A (Cas A). 
We find that, for a large region of the axion parameter space, the gamma ray flux from 
the directions of SN 1987A and Cas A should be easily observable by the Fermi Large Area Telescope (Fermi-LAT).  
Observations of this signal would provide information on the properties of these particles. 
The lack of such observations allows us to place limits on the HALP-photon coupling (or equivalently, $f_a$) 
as a function of the HALP mass.  
We find that for  $\fa=10^{12}\,\GeV$, the HALP mass is constrained in the range from $100\,\eV$ to $1 \,\GeV$.  
The most stringent constraint on the Peccei-Quinn scale applies to HALP masses 
of approximately 100~MeV and restricts $\fa$ to values greater than about $10^{15}\,\GeV$. 

This paper is organized as follows.
In the next section, we analyze the HALP production from the SN core, 
calculate the corresponding photon flux from their decay, and compare our results to the Fermi-LAT observations, 
thereby obtaining limits for HALP parameters. 
%will be described in section \ref{sec:ConstraintsFormFermiTelescope}.
In section \ref{sec:Conclusions}, we review and discuss our results.
Finally, in the appendix we provide a quantitative analysis of the effect of a lepton decay channel, neglected in the rest of the paper.

%---SECTION 3: RESULTS OF ANALYSIS (PLUS SOME OTHER STUFF)-------------------------------------
\section{Heavy Axion-like Particles and Supernovae}
\label{sec:HeavyAxionsAndSupernovae}

% THERE ARE 3 MAIN IDEAS IN THIS SECTION:
% (1) outline of what we are doing
% (2) Equations of how this is done
% (3) Results from the equations.

%Supernovae provide an excellent environment for our purpose, for two reasons: \textit{a)} the axion can be efficiently produced; \textit{b)} the distance from the Earth is large enough to allow axions to decay into photons. The resulting photon flux for $\fa=10^{12}\GeV$ can indeed be rather big, compared to the sensitivity of FERMI LAT, see Fig. \ref{fig:graph_1}, providing a constraint on the axion mass up to about $1\GeV$.

%---INTRODUCTION TO WHAT WE ARE DOING---------------------------

In this section, we will first calculate the HALP luminosity from SNe, disregarding relation~(\ref{mass}), and then the flux of photons that we could expect from HALP decay on Earth.  Combined, these calculations allow us to derive new limits on HALP properties, based on the possible detection of the emitted photons from HALP decay.  
%Our calculation of the luminosity, in section~\ref{sec:GeneralizationOfTheEnergyLossMechanism}, will also allow us to generalize the standard SN energy loss argument to include a massive pseudoscalar, disregarding relation~(\ref{mass}). Following this, we discuss in detail our new limits in section~\ref{sec:ConstraintsFromDecayIntoPhotons}.

\subsection{Generalization of the energy loss mechanism}
\label{sec:GeneralizationOfTheEnergyLossMechanism}

Massive pseudoscalars may be produced efficiently in a SN core, due to the high temperature and density.  
With temperatures of $T\sim 30\,\MeV$ and densities of $\rho \sim 3\times 10^{14}~{\rm g~ cm}^{-3}$, 
the most efficient axion production mechanism in a SN core
is the nucleon-nucleon bremsstrahlung process~\cite{Raffelt:1996wa}
%~\cite{Iwamoto:1984ir,Brinkmann:1988vi,Raffelt:1993ix,Hanhart:2000ae,Giannotti:2005tn,Turner:1987by,Burrows:1988ah},
%
\begin{equation}\label{brem}
	N~N\to N~N~a~,
\end{equation}
where $N$ is a nucleon.   
The bremsstrahlung production of massive axions has been studied in ref.~\cite{Giannotti:2005tn}, using the one pion exchange approximation.  
 We use the same formalism to obtain our HALP emission rates.  
%We show here how we calculated the gamma-ray flux on Earth from the decay of massive axions produced in SN.  
Setting $C_N\simeq 1$ to normalize the HALP-nuclear coupling, and a SN core of radius of $10$~km, we obtain the production rate of massive pseudoscalar in SN, 
%number of axions with energy between $\omega$ and $\omega+d\omega$ produced per second in the SN core is~\cite{Nesti} 
%
\begin{equation}\label{eq:emission_number}
	\frac{dN}{d\omega}\simeq \frac{1.25\times 10^{49}\,\left(\MeV\,\s\right)^{-1}}{\f^2} 
	 \left(\frac{T}{30\;\MeV}\right)^{3/2}\left(\frac{\rho}{3\times 10^{14}\,\den}\right)^{2}
	 \beta^3 \phi \left(\frac{\omega}{2T}\right)\,,
\end{equation}
where $dN$ is the number of axions with energy between $\omega$ and $\omega+d\omega$,
%$T$ and $\rho$ are the SN temperature and density respectively, 
$\beta$ is the axion velocity as a fraction of the speed of light, and $\f=\fa/10^{12}\GeV$.  The function $\phi(x)$ is given by 
\begin{eqnarray}
\label{phi}
	\phi(x)= \frac52\,x^2\,e^{-x}\, K_1(x)\,,
\end{eqnarray}
where $K_1$ is the modified Bessel function of the second kind. 
The normalization of $\phi$ is such that $\int_0^\infty \phi\,dx=1$.
In the relativistic case ($\beta\simeq1$), all the energy dependence is contained 
in $\phi(x)$ which then represents the axion spectrum.  
The maximum of $\phi(x)$ occurs at $x\simeq 0.62$, approximately corresponding to an energy of $1.2 \,T$.  At high energies,
%, the axion production rate is exponentially suppressed
%
\begin{equation}
	\phi \left(\frac{\omega}{2T}\right) \simeq \frac52\sqrt{\frac{\pi}{2}}\,
	\left(\frac{\omega}{2T}\right)^{3/2}e^{-\omega/T}\,,\qquad (\omega\gg T)
	\label{eq:phihit}
\end{equation}
%B
and thus, for $m_a\gg T$, the production rate is suppressed by an exponential factor.  
This is the origin of the Boltzmann suppression factor expected for the production  of heavy particles in SNe.   
Additionally, in the nonrelativistic regime, the production rate of Eq.~(\ref{eq:emission_number}) is suppressed 
by the velocity factor, $\beta^3$.  

%, which emerges in the calculation of the 
%S matrix for the bremsstrahlung process (\ref{brem}), in the case of emission of an axion with finite mass~\cite{Giannotti:2005tn}. 

As we are interested in large values of the PQ constant, corresponding to weak coupling, we assume the HALPs free stream from the SN core and decay into photons only when they are free of the SN matter.  In the free-streaming regime, the total energy loss rate via HALPs, or HALP luminosity, is
%for a general axion-like particle 
%
\begin{equation}\label{luminosity}
	L_a=\int_{m_a}^\infty \omega \,\left(\frac{d N}{d\omega}\right)d\omega\,,
\end{equation}
with $dN/d\omega$ given by Eq.~(\ref{eq:emission_number}).

Now we can easily generalize the novel particle cooling limit of SN 1987A, originally derived for a massless axion, to a general HALP.  
Requiring that the novel particle luminosity of Eq.~(\ref{luminosity}), evaluated at typical core condition 1 s after the collapse
($T\sim 30\,\MeV$, $\rho \sim 3\times 10^{14}~{\rm g~ cm}^{-3})$  
does not exceed $3\times 10^{52}$ erg/s, we find the mass-dependent constraint on the PQ constant
shown in Fig. \ref{fig:Contour1}.  Our result is in agreement with the standard bound, constraining $f_a$ to be greater than $10^9$~GeV for masses less than 30 MeV.  As expected, the bound is strongly reduced for larger masses.

%
%From the duration of the neutrino signal, it is possible to show that the energy loss rate (per gram) through any novel particle has to be less than $10^{19}{\rm erg\,g^{-1}}\,{\rm cm}^{-3}$~\cite{Raffelt:1996wa}.
%
\FIGURE{
	\epsfig{file=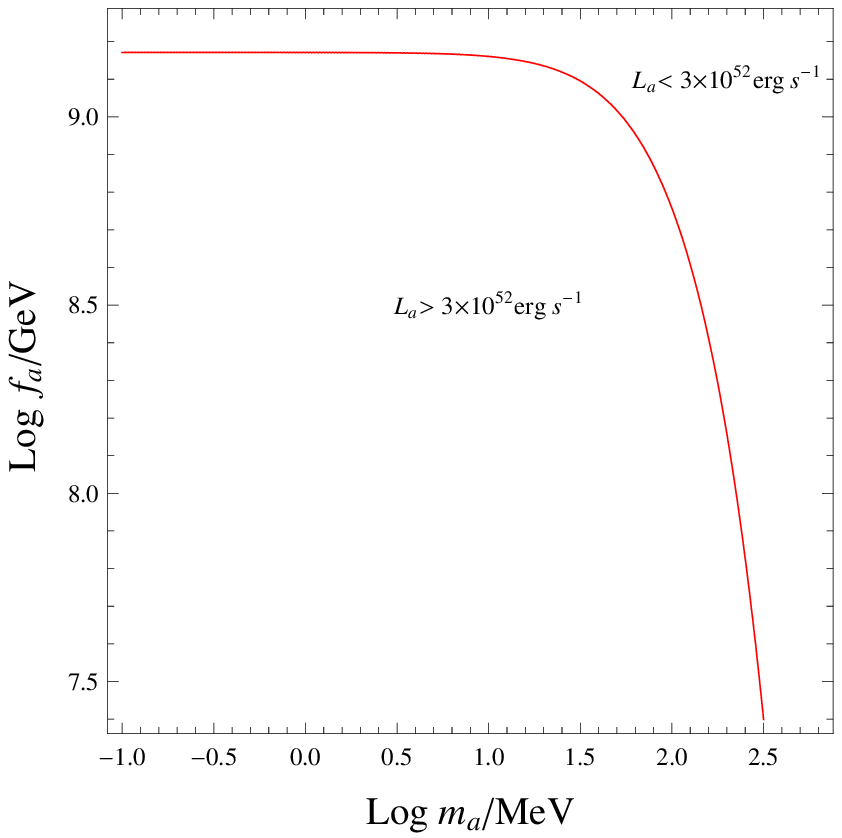,width=0.7\textwidth}
	\caption{Excluded region for the PQ constant as a function of the HALP mass, based on novel particle cooling in SN 1987A.}
	\label{fig:Contour1}
	}

\subsection{Constraints from decay into photons}
\label{sec:ConstraintsFromDecayIntoPhotons}
In the rest of this section, we discuss the stronger phenomenological constraint on axion-like particles which arises from their subsequent decay into photons.  Weakly coupled particles will freely stream out of the SN core, and, if massive enough, decay into photons before reaching Earth, generating an observable photon flux.  This flux would be observable for a long time after the SN explosion, as massive particles propagate more slowly than neutrinos and photons produced directly in the SN, and their decay time is statistically distributed. Therefore, a flux of gamma photons would still be observable today.

We compare the predicted photon flux from decaying HALPs with the observations of gamma-rays from the Fermi Large Area Telescope, a detector designed to detect high-energy gamma-rays up to $300\GeV$, 
with an angular resolution of $1^\circ$ or so at energies of a few $100\MeV$ \cite{Atwood:2009ez}.  

We study both SN 1987A and Cas A. 
We are interested in the photon signal from the decaying-HALP observable today, after a time of some $10^{9}\s$ from the direct observation of the SN explosion. As the SNe distances are of the order $10^{12}\s$, decay-produced photons coming from an angle larger than $\sim 1^\circ$ will not have had time to reach Earth.  Therefore, both sources would appear as point sources, if observed by Fermi-LAT.

%As discussed in more detail later, the delay time between the direct photon or neutrino signal and the decaying-HALP photon signal is of order $10^{9}\s$. As the SNe distances are of the order $10^{12}\s$, decay-produced photons coming from an angle larger than $\sim 1^\circ$ will not have had time to reach Earth.  Therefore, both sources would appear as point sources, if observed by Fermi-LAT.  

Cas A, which seems to correspond to the Flamsteed's SN of 1680, 
is the closest known core-collapse SN~\cite{Pavlov:1999et}
from which a flux of gamma rays in the range we are interested in has not been observed~\cite{Hannestad:2001xi}.  
Recently observed gamma rays from Cas A~\cite{Funk:2010cg} 
are too energetic to be caused by the mechanism that we are describing here.  No gamma-ray counterpart of SN 1987A has yet been detected.  As they are not observed at the energies we are interested in, we compare our predicted fluxes with the point source sensitivity of the Fermi-LAT, $< 6 \times 10^{-9} {\rm photons\,cm}^{-2} {\rm s}^{-1}$ for photons of energies $\omega_\gamma \geq 100$~MeV, at the latitudes of the SNe we are considering.  
As we shall see below, the analysis from Cas A will provide the most stringent bounds on the axion-photon coupling, because of its closer distance to Earth.

%--END OF THINGS IN THE INTRODUCTION--------

%---SECTION OF EQUATIONS--------------

We now compute the gamma-ray flux on Earth from the decay of massive pseudoscalar particles produced in SNe.  
In the following, we assume that the HALP couples only to photons and nucleons.  
The definitions of these coupling strengths can be found after Eq.~(\ref{a-couplings}). 

%Fig.~\ref{fig:Contour1} shows the region where the axion luminosity (\ref{luminosity}),
%evaluated at typical core condition 1 s after the collapse
%($T\sim 30\,\MeV$, $\rho \sim 3\times 10^{14}~{\rm g~ cm}^{-3})$  
%does not exceed $3\times 10^{52}$ erg/s.

%\begin{figure}[t]
%	\centering
%		\includegraphics[width=0.30\textwidth]{phi}
%	\caption{phi}
%	\label{fig:graph_phi}
%\end{figure}

%\footnote{In general, this hypothesis is not a completely justified since quantum corrections would induce a pseudoscalar-lepton coupling in the effective Lagrangian through the interaction with photons. Thus, massive axions could have another decay channel. This effect would therefore reduce the expected photon flux and relax somewhat our bounds. This is, however, not a very significant effect and will be neglected here.}. 
The HALP lifetime is then given by
\begin{equation}
\label{tau}
	\tau\equiv  \tau_{a\to \gamma \gamma}
	\simeq 1.7\times 10^{17}\, \f^2\, \left(\frac{m_a}{\MeV}\right)^{-3}\, {\rm s}\,.
\end{equation}
SN 1987A occurred at a distance of $\R\simeq 51.4$~kpc from Earth or approximately $5.3\times 10^{12}$~s in natural units ($c=1$).  Cas A is located $\R\simeq 3.4$~kpc or $3.5\times 10^{11}$~s from Earth.  Comparing these numbers with the lifetime of Eq.~(\ref{tau}), we see that a sizable fraction of HALPs produced can decay into photons before reaching Earth if their mass is large enough.  The actual number of HALPs that decay into photons after a time $t_a$ from production is 
\begin{equation}\label{eq:axion_decay}
	dN_{a\to\gamma\gamma}=\left[1-\exp{\left(-\frac{t_a}{\gamma\tau}\right)}\right]dN\,,
\end{equation}
where $\gamma=\omega/m_a$ is the HALP Lorentz factor and $\omega$ is the HALP energy.

%Quantitatively, if we indicate with $\beta t_a$ the distance that the axion goes before decaying ($\beta$ is the axion velocity), and with with $R_\gamma=t_\gamma$ the distance that the produced photon (from the axion decay) goes to reach the earth, then 
%\begin{equation}
%	R_a^2+R_\gamma^2+2R_a R_\gamma \cos{\theta}=R_{\rm SN}^2
%\end{equation}

In order to calculate the expected photon flux today, we need to find the HALP decay time $t_a$ in Eq. (\ref{eq:axion_decay}).  
In natural units, the distance traversed by the HALP before decay, $\beta t_a$, and that by the photons produced by HALP decay which travel to Earth, $t_\gamma$, are geometrically related to the SN distance, $\R$, via
\begin{equation}
	(\beta t_a)^2+t_\gamma^2+2(\beta t_a) t_\gamma \cos{\theta}=d_{\rm SN}^2 \,,
\end{equation}
where $\theta$ is the photon emission angle.  This can be solved for $t_a$ in terms of known parameters,

 %
 %through the equation
%%%
%%\begin{equation}
%%\label{ta}
%%	\Delta t=t_a \left[1-x\beta+\sqrt{\left(\frac{t_0}{t_a}\right)^2+\beta^2(x^2-1)}\,\,\right]-t_0
%%\end{equation}
%%%
%%
%\begin{equation}
%\label{ta}
%	\frac{\Delta t+t_0}{t_a }=1-x\beta+\sqrt{\left(\frac{t_0}{t_a}\right)^2+\beta^2(x^2-1)}
%\end{equation}
%%
%where $x=\cos{\theta}$. This can be easily solved for $t_a$
%
\begin{equation}\label{eq:ta}
t_a=\frac{\left(d_{\rm SN}+\Delta t\right)\left(1-x\beta\right)
-\sqrt{
d_{\rm SN}^2\left(1-x\beta\right)^2
+\beta^2\Delta t\left(2d_{\rm SN}+\Delta t\right)\left(x^2-1\right)}}
{1-2x\beta+\beta^2} \,,
\end{equation}
where $x=\cos{\theta}$ and $\Delta t=t_a+t_\gamma-\R$ is the HALP \textit{delay time}.  
This is the time between the signal seen on Earth from a photon directly produced in the SN and that 
from the photons produced via HALP decays.  In this analysis, the delay time is that between direct observation of each SN and present day.  For SN 1987A, $\Delta t\simeq 22$ years or $6.9\times10^{8}$~s and for Cas A, $\Delta t\simeq 320$~years or $1.0\times10^{10}$~s.

Notice that for both SN 1987A and Cas A, $\Delta t\ll \R$. 
For the purpose of our calculation, we can therefore ignore the terms of order $(\Delta t/\R)^2$ or higher, 
which contribute less than $1\%$ to the time $t_a$. With this simplification $t_a$ becomes: 
\begin{equation}\label{eq:ta2}
t_a\simeq\frac{\Delta t}{1-x\beta} +O\left(\frac{\Delta t}{\R}\right)^2 \,.
\end{equation}
%

%The approximation of forward emission %$(x\to 1)$
%%%, \textbf{discussed in Proceedings}, 
%%is particularly simple 
%%
%\begin{equation}\label{eq:SIMP_ta}
%\lim_{x\to 1} t_a=\frac{\Delta t}
%{1-\beta}
%\end{equation}
%%
%%in the limit of forward emission ($x\to 1$).
%represents an enormous simplification which, as we shall see later, allows some analytical estimates of the expected flux, 
%and represents often a good approximation. %especially in the high mass limit and ultrarelativistic limit . 
%We will however present the numerical result for the case of general emission angle.

We now proceed to calculate the resulting photon spectrum from HALPs decay.  
Here, we will make the simplifying assumption that the two photons are emitted 
with the same energy, $\omega_\gamma=\omega/2$.
This assumption is reasonable in both the nonrelativistic and the ultrarelativistic limits, and 
numerical tests show that the effect of this approximation on 
the calculation of the photon flux is small.  The resulting photon spectrum is given by
\begin{equation}
\label{g_spectrum}
	\frac{dN_{\gamma}}{d\omega_\gamma}=
	4 \int_{-1}^{1} \left[1-\exp{\left(-\frac{t_a}{\gamma\tau}\right)}\right]\frac{dN}{d\omega} f(x) \,dx \,,
\end{equation}
where 
\begin{equation}\label{eq:angle}
	f(x)=\frac12 \frac{1-\beta^2}{(1-\beta x)^2}
\end{equation}	
is the normalized probability distribution for the photon emission angle~\cite{Landau}. 

The expected HALP-produced photon flux on Earth can now be calculated from Eqs. (\ref{eq:ta}), (\ref{g_spectrum}), and  (\ref{eq:angle}) 
in terms of the axion mass, the delay time $\Delta t$, and the SN distance $\R$.  Assuming isotropic emission, massive axion-like particles would produce a flux of high energy photons, 
\begin{equation}
\label{g_flux}
	\frac{dF_{\gamma}}{d\omega_\gamma}=\frac1{4\pi d_{\rm SN}^2}\frac{dN_{\gamma}}{d\omega_\gamma}\,.
\end{equation}
%
%Since the production mechanism is (partially) thermal, we expect the photon spectrum to be picked at energies of the order of the SN core temperature. 
%An example of spectrum, from SN 1987A axions of mass 1 MeV and PQ constant $10^{12}$ GeV, is shown in Fig.~\ref{fig:spectrum}.
%%So, our results should be compared to the present experimental observations of gamma rays. 

%\FIGURE{
%	\epsfig{file=Spectrum1,width=0.35\textwidth}
%%	\epsfig{file=Spectrum2,width=0.35\textwidth}
%	\caption{Photon spectrum from SN 1987A axions. (Left panel) axion mass $m_a=1$ MeV. (Right panel): axion mass $m_a=300$ MeV. 
%	In both cases $\fa=10^{12}~\GeV$.}
%	\label{fig:spectrum}
%	}
%

In order to compare our prediction with the Fermi-Lat data, we need to calculate the 
expected number of photons with energy greater that $100~\MeV$ generated by the HALPs decay,
\begin{equation}\label{Eq_F100}
	F_{100}=\int_{100\MeV}^{\infty}\left(\frac{dF_{\gamma}}{d\omega_\gamma}\right)\,d\omega_\gamma \,,
\end{equation}
as a function of the HALP mass, $m_a$, and the PQ constant, $\fa$.

%Our numerical result is shown in figure \ref{fig:graph_1} and \ref{fig:graph_2}.
%However, some analytical estimates are possible and useful to understand the behavior of $F_{100}$.

Notice that since we are only interested in photons with energy greater than $100$ MeV, 
we can use the approximation (\ref{eq:phihit}) in Eq. (\ref{Eq_F100}). Setting $T=30$MeV and $\rho=3\times 10^{14}\,\den$, 
and defining $\omega_1=\omega_\gamma/1\MeV$ and $d_{\rm 87A}=51.4$ kpc,
we find
\begin{eqnarray}\label{Eq_F100_2}
	  F_{100}\simeq \frac{3\,\f^{-2}}{{\rm cm}^2\,\s}  \left(\frac{d_{\rm 87A}}{\R}\right)^2   %\nonumber \\
%	 &&~~~~\times  
%	\left[1-\exp{\left(-\frac{t_a}{\gamma\tau}\right)}\right]   \beta^3 \omega_1^{3/2}e^{-\frac{\omega_1}{15}}f(x)dx d\omega_1
	 \int_{100}^{\infty}   \beta^3  \omega_1^{3/2}e^{-\omega_1/15} \,d\omega_1
		 \int_{-1}^{1} \left[1-e^{-t_a/\gamma\tau}\right]   f(x)dx   \,.
\end{eqnarray}
Using Eq. (\ref{eq:ta2}) for the axion decay time, 
the angular integral can be easily estimated in both the relativistic and nonrelativistic regimes.

In the relativistic regime (which applies to axions with mass $m_a\ll 200$ MeV, since we are interested only in photons of energy above 100 MeV) 
the calculation of $F_{100}$ can be analytically performed to the end, with the use of very reasonable approximations.
In this regime $t_a/\gamma\tau \sim \Delta t/\gamma\tau\ll 1$, 
and so the term in the square brackets in Eq. (\ref{Eq_F100_2})
can be substituted with $t_a/\gamma\tau$. 
In addition, we can neglect the $\beta^3\simeq1$ factor in Eq. (\ref{Eq_F100_2}). 
With these approximations, the integral in $dx$ becomes straightforward, and Eq. (\ref{Eq_F100_2}) reduces to
\begin{eqnarray}\label{Eq_F100_3}
%	  && F_{100}\simeq \frac{3\,\f^{-2}}{{\rm cm}^2\,\s}  \left(\frac{d_{\rm 87A}}{\R}\right)^2   
%	 \int_{100}^{\infty}   \left(\gamma\tau\right)^{-1}  \omega_1^{3/2}e^{-\omega_1/15} \,d\omega_1  
%		 \int_{-1}^{1} t_a f(x)dx   \nonumber \\
%		 && \frac{2.4\,\f^{-2}}{{\rm cm}^2\,\s}  \left(\frac{d_{\rm 87A}}{\R}\right)^2 
%		 \int_{100}^{\infty}   \left(\frac{\gamma\Delta t}{\tau}\right)   \omega_1^{3/2}e^{-\omega_1/15} \,d\omega_1= \nonumber \\
%		 && \frac{2\times 10^{-8}\,m_1^2\,\f^{-4}}{{\rm cm}^2\,\s}  \left(\frac{d_{\rm 87A}}{\R}\right)^2\left(\frac{\Delta t}{d_{\rm 87A}}\right)
%		 \int_{100}^{\infty}   \omega_1^{5/2}e^{-\omega_1/15} \,d\omega_1
%		 = \nonumber \\
%		 && 
		 F_{100}\simeq \frac{6.8 \times 10^{-5}}{{\rm cm}^2\,\s}  
		 \left(\frac{d_{\rm 87A}}{\R}\right)^2\left(\frac{\Delta t}{\Delta t_{\rm 87A}}\right)\,\left(\frac{m_a}{1\MeV}\right)^2\,\f^{-4}\,,
		   \quad (m_a\ll 200\,\MeV),
\end{eqnarray}
where $\Delta t_{\rm 87A}=22$ yr.

For axions heavier than $200$MeV, the lower integration limit in the photon energy 
in Eq. (\ref{Eq_F100_2}) should be substituted with $2m_a$. 
Since this is much larger than the SN core temperature, the exponential in the integral dominates, 
and so the integral is peaked in the region of low energy. 
In this region, the exponential term is small, $\exp[-t_a/\gamma\tau]\ll 1$, and so 
\begin{eqnarray}\label{eq_heavy}
	  F_{100}\simeq \frac{3\,\f^{-2}}{{\rm cm}^2\,\s}  \left(\frac{d_{\rm 87A}}{\R}\right)^2   %\nonumber \\
%	 &&~~~~\times  
%	\left[1-\exp{\left(-\frac{t_a}{\gamma\tau}\right)}\right]   \beta^3 \omega_1^{3/2}e^{-\frac{\omega_1}{15}}f(x)dx d\omega_1
	 \int_{\frac{m_a}{2{\tiny \MeV}}}^{\infty}   \beta^3  \omega_1^{3/2}e^{-\omega_1/15} \,d\omega_1
	 \,,  \qquad (m_a\gg 200\MeV)\,.
\end{eqnarray}
The above expression %Eq. (\ref{eq_heavy}) 
cannot be explicitly calculated because of the $\beta^3$ factor.
However, since the integral has to be evaluated in the region where the exponential dominates, 
$\omega_\gamma\geq m_a\gg 15\MeV$, the expression for $F_{100}$ is going to behave as
%the typical form of these integrals in the region where the exponential dominates, $\omega_1>m_1\gg 15$, is
\begin{equation}
	F_{100}\sim \f^{-2}\, e^{-m_a/\zeta} \,,  \qquad (m_a\gg 200\,\MeV),
\end{equation}
with $\zeta\sim 30\MeV$, predicting an exponential suppression of the photon flux for large axion mass.

%The results of the numerical integration are
%shown in Figs.~\ref{fig:graph_1} and \ref{fig:graph_2}.
%%
%\DOUBLEFIGURE[]{Flux100}{Contour}
%%\label{fig:graph_1}
%{Flux of photons (from axions decay) with energy $\omega_\gamma>100\MeV$ expected from SN Cas A  (blue curve on top) and SN1987A (red curve on the bottom). In both cases $\fa=10^{12}\GeV$.}
%	{Exclusion plot for axion models with mass up to $\sim 1$GeV, and $\fa>10^{11}\GeV$. The exclusion region shows the axion parameters which would cause a photon flux $F_{\gamma}> 6 \times 10^{-9} {\rm cm}^{-2} {\rm s}^{-1}$.}
%\FIGURE{
%	\epsfig{file=Flux100,width=0.4\textwidth}
%	\epsfig{file=Contour,width=0.35\textwidth}
%	\caption{(Left panel): Flux of photons (from axions decay) with energy $\omega_\gamma>100~\MeV$ expected from SN Cas A  (blue curve on top) and SN1987A (red curve on the bottom). The PQ constant is $\fa=10^{12}~\GeV$ in both cases.
%	(Right panel): Exclusion plot for axion models with mass up to $\sim 1$GeV, and PQ constant $\fa>10^{11}~\GeV$. 
%	The exclusion region shows the axion parameters which would cause a photon flux $F_{100}> 6 \times 10^{-9} ~{\rm cm}^{-2} {\rm s}^{-1}$.}
%	\label{fig:graph_1}
%	}
%	
	\FIGURE{
	\epsfig{file=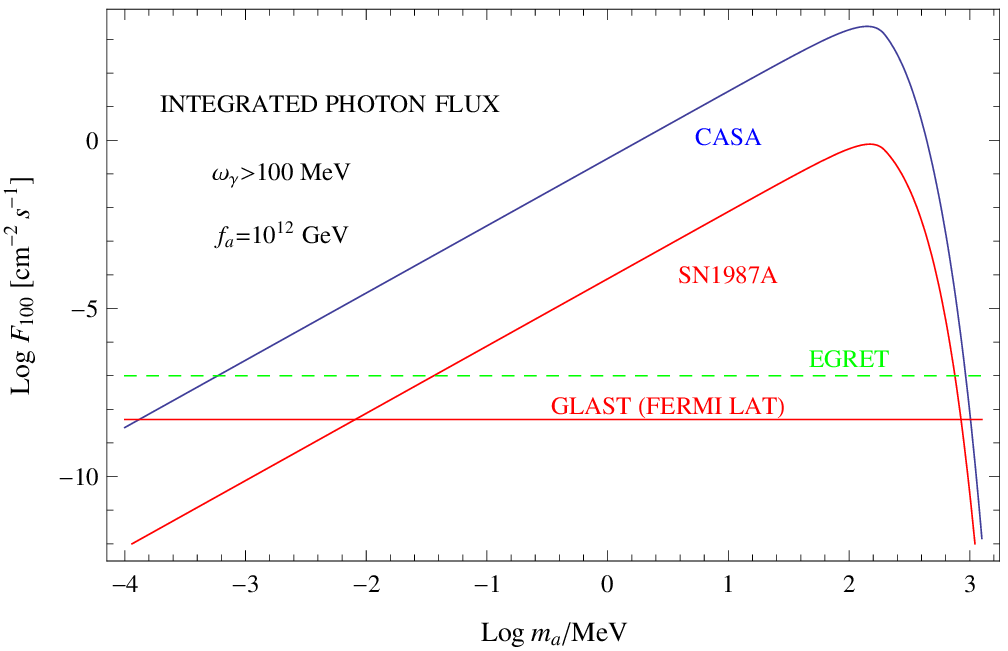,width=0.7\textwidth}
	\caption{Flux of photons, $F_{100}$, from HALP decay with energy greater than $100~\MeV$ expected from SN Cas A  (blue curve on top) and SN 1987A (red curve on the bottom) as a function of HALP mass. The PQ constant is $\fa=10^{12}~\GeV$ in both cases.  The point source sensitivity of the Fermi-LAT and earlier gamma-ray telescope, EGRET, are also shown.}
	\label{fig:graph_1}
	}

	\FIGURE{
	\epsfig{file=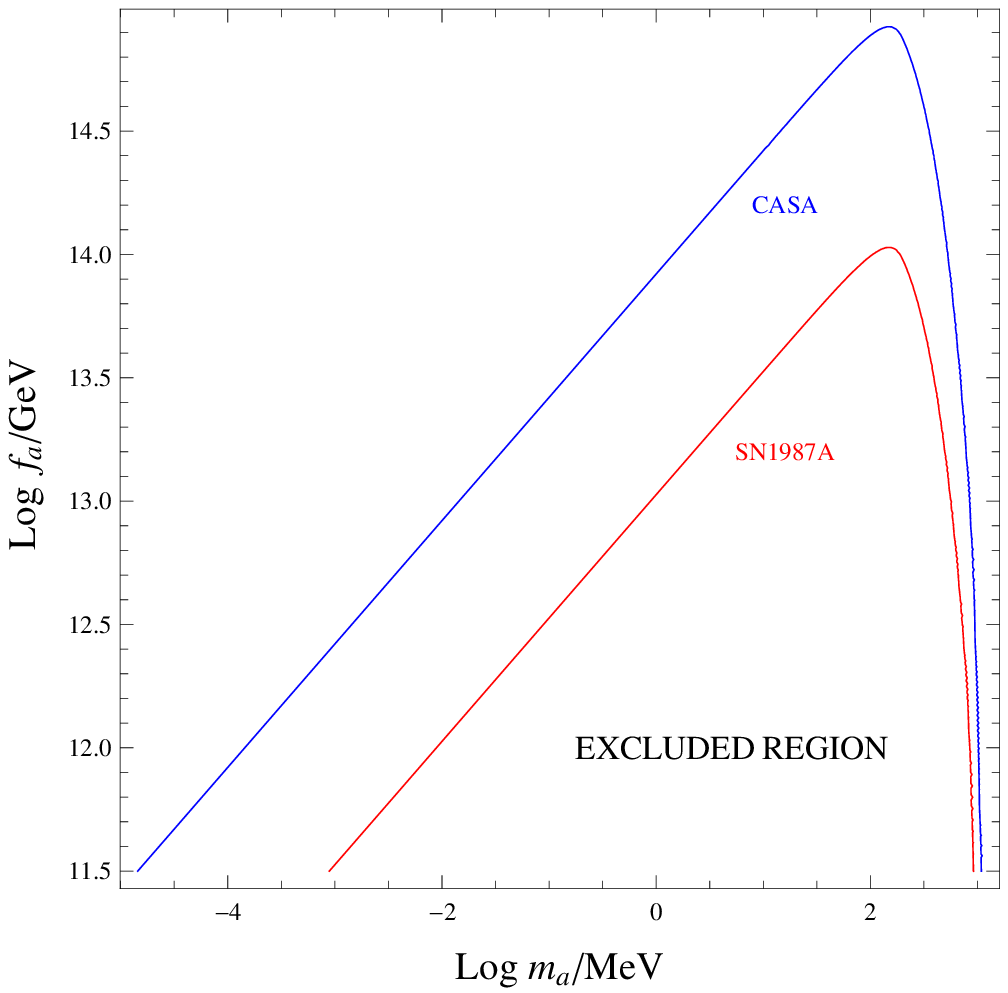,width=0.65\textwidth}
	\caption{Exclusion plot for HALP models with mass up to 1~GeV, and PQ constant $\fa>10^{11}~\GeV$. 
	The exclusion region shows the axion parameters which would cause a photon flux greater than $6 \times 10^{-9} ~{\rm cm}^{-2} {\rm s}^{-1}$. }
	\label{fig:graph_2}
	}

We show the results of the numerical integration of Eq. (\ref{Eq_F100}) in Fig.~\ref{fig:graph_1} and \ref{fig:graph_2}.
In Fig.~\ref{fig:graph_1}, we compare the photon flux from HALP decay, $F_{100}$, with the Fermi-LAT point source sensitivity
as a function of the HALP mass, for the case of $\fa=10^{12}\GeV$.
For comparison, we also show the detection limit from the Energetic Gamma Ray Experiment Telescope (EGRET), 
an older gamma-ray telescope with point source sensitivity of approximately $10^{-7} {\rm photons\,cm}^{-2} {\rm s}^{-1}$, 
for photon energies of $\omega_\gamma \geq 100$~MeV, at high latitudes~\cite{Sreekumar:1997un}.  Where the predictions are above the Fermi-LAT point source sensitivity, we are able to provide an exclusion limit.  In this case, the region of HALP mass between $10\,\keV$ and $1\,\GeV $ is excluded from the lack of observation of a photon flux from SN 1987A. The data from Cas A exclude a larger mass region, from approximately $100\,\eV$ to $1\GeV$.

Fig.~\ref{fig:graph_2} shows the complete exclusion region obtained numerically, in the  
$m_a$-$\fa$ plane.  The excluded values are those for which Fermi would be able to observe our predicted spectrum of gamma-rays.
The constraint from Cas A is more stringent, due to both a smaller value of $\R$ 
and a larger value of $\Delta t$, with respect to SN 1987A (cfr. Eq. (\ref{eq_rel_approx}) below).

The agreement with the analytical approximation in the relativistic regime is within a few percent, 
and the exponential suppression in the very massive regime is also evident.

The contour line in the low mass region ($m_a\lesssim 100\MeV$) shown in Fig.~\ref{fig:graph_2} is, according to Eq. (\ref{Eq_F100_3}), 
a straight line with equation
%with Fermi sensitivity, $F_{100}=6 \times 10^{-9} ~{\rm cm}^{-2} {\rm s}^{-1}$, 
%we find the equation for the line shown in figure \ref{fig:graph_2} for $m_a\lesssim 100$ MeV:
\begin{equation}\label{eq_rel_approx}
	\log \left(\frac{f_a}{1\GeV}\right)\simeq 13 + 
	\frac14\log \left[\left(\frac{d_{87A}}{\R}\right)^2 \left(\frac{\Delta t}{\Delta t_{\rm 87A}}\right)\right]
	+\frac12\log \left(\frac{m_a}{1\MeV}\right)\,.  %\qquad ({\rm relativistic~ approximation}),
\end{equation}
The above relation gives a quantitative prediction of the effect of the SN distance and the delay time 
on the expected photon flux.
%which predicts the correct slope and the correct distance of the lines for SN 1987A and Cas A shown in figure \ref{fig:graph_2}.

From the figures, we see that the bounds on $\fa$ are peaked around the value $m_a\sim 100$~MeV, with $\fa$ restricted to being greater than approximately $10^{15}$~GeV.  
The bounds are strongly suppressed at lower and higher masses.  
This is not surprising, as the the HALP flux is maximized if the HALP can be easily produced, 
but is also heavy enough to efficiently decay.
This differs from the standard energy loss (or novel particle cooling) arguments, which are particularly efficient for low mass particles.

\section{Discussion and Summary}
\label{sec:Conclusions}

Here we briefly summarize and comment on the key points of our analysis and discuss possible future directions.  
The main motivation for this work is the analysis of the heavy-axion-like particles (HALP) parameter space, 
particularly in the case when the mass and PQ constant are not directly related.  
%Our analysis provides new constraints in the case of large mass, specifically in the range between approximately $100\,\eV$ and 1~GeV.

We have calculated the production rate of massive HALPs in a SN core and analyzed the 
photon flux expected from their decay into photons. 
The flux has then been compared to the results of the Fermi Large Area Telescope. 
Our main results are shown in Figs.~\ref{fig:graph_1} and \ref{fig:graph_2}.  
For a HALP mass of $\sim 100\,$MeV, we find that $\fa$ must be greater than roughly $10^{15}\,\GeV$.  
For the specific case of $\fa=10^{12}\,\GeV$, we find that the values of the HALP mass between approximately $100\,\eV$ and $1 \,\GeV$ are excluded.  

The idea of constraining radiatively decaying axions from gamma-ray observations of 
SN 1987 A was considered already shortly after the SN 1987 A event (see, e.g., \cite{Kolb:1988pe}). 
The original analysis was applied, however, only to photons from light axions decay, 
which reach the earth within a few seconds from the observation of the explosion.
%The original approach, however, applies only to photons from light axions decay, which 
%reach the earth within a few seconds from the observation of the explosion. 
%In this work, we generalized the argument to make it applicable to later observations and, consequently, to other SN events.

In general, the axion bounds on $\fa$ from stellar cooling are constant for masses up to a few $T$, 
where $T$ is the relevant stellar temperature, and are exponentially suppressed for higher masses.
Our bound, instead, is maximal for HALP masses around $100$~MeV (a few times the SN core temperature of 30~MeV), 
but is reduced in the high and the low mass region.
In fact, in the mechanism we have studied, two different effects play a role:  HALP production in the SN core, and HALP decay into photons. 
The production rate of heavy HALPs is exponentially suppressed as the energy required to produce HALPs increases, but the decay rate into photons increases with the third power of the mass.

We have also investigated the SN 1987A novel cooling limit that applies when the HALP mass is independent of the PQ constant. 
The standard energy loss argument from the analysis of the observed SN 1987A neutrino signal becomes inefficient for axion masses greater than the SN core temperature, approximately $30$ MeV. This is quantitatively shown in Fig.~\ref{fig:Contour1}.  This limit is in agreement with previous bounds derived from SN 1987A for the standard axion. 

In deriving these new bounds, we have made a number of hypotheses.  
First, and most importantly, we have calculated the axion production rate in the one-pion-exchange (OPE) approximation, which is not completely justifiable.
As shown in \cite{Hanhart:2000ae}, this approximation might induce an overestimate of the production rate.
Although this conclusion cannot be directly applied to the results in our paper,
since the analysis in \cite{Hanhart:2000ae} does not take into account the effects of finite axion mass,
and although our bound on $\fa$ depends only on the fourth-root of the production rate,
it is evident that our results should be considered within some not easily quantifiable uncertainty, 
as many details of the SN physics and particles production in its core are yet not completely known 
(see also the discussion in \cite{Raffelt:1996wa}, page 120).

We have also assumed that our axion-like particle interacts with nucleons and photons only. 
We have not included the possibility of axion decay into leptons,	even when this would be kinematically possible ($m_a>2m_\ell$). 
	Even if one assumes that the pseudoscalar-lepton coupling is absent in the Lagrangian (e.g. following hadronic axion models), 
	the coupling with photons would necessarily induce a (suppressed) coupling with electrons, and so massive axions 
	could have another decay channel.  This effect would reduce the expected photon flux and somewhat relax our bounds. 
	Our numerical tests discussed, in the appendix, show, however, that this would not be a very significant effect for large PQ constant.   
%	\item Our analysis shows that in some cases massive axions would be allowed only in the anthropic region.
	
Finally, we have not considered the HALP couplings to nucleons and photons as independent.  
These couplings are still related in the same manner as standard axion models, via the PQ constant. 
This may not be the case for a more generic pseudoscalar.  In principle, it is possible to tune the 
axion-photon coupling to very small values without changing the axion coupling with nuclei~\cite{Kaplan:1985dv}. 	
This would, of course, modify our bounds.
	
Other possible arguments could be used to constrain the region of the parameter space which we have analyzed.
For example, for large PQ constant and small mass, 
axions would have a lifetime longer than the age of the universe. 
Therefore, another possible constraint could be derived from the analysis of the axion thermal production, 
and the calculation of the axion relic abundance today. 
A full calculation of the HALP relic abundance, based on references~\cite{Turner:1986tb,Masso:2002np}, shows that this is the case for $f_a\lesssim 5\times 10^{12}\GeV$, and $m_a>$ a few $10^{-4}\MeV$, providing that the HALP lifetime exceeds the age of the universe. 
Therefore, from our analysis, it results that the SN argument which we used in this work is more efficient in constraining the region of high $\fa$. A full analysis of other possible constraints in that region would, however, be very interesting and should be considered in the  future.

%However, the relic abundance expected from axions in the region we explore, $\fa\gtrsim 10^{12}\GeV$, would be negligible~\cite{Turner:1986tb,Masso:2002np}. 

%Our results are comparable to the HALP bound previously obtained by Masso and Toldra \cite{Masso:1997ru}.  
%This bound can be interpreted as limiting $f_a$ to be greater than $10^{10}$~GeV for ALPs of mass 1~keV to 1~GeV, in the same context as this paper .  As per our study, this bound is based on the assumption that the only ALP decay channel is to photons.  Our limits, shown in Fig.~\ref{fig:graph_2}, are applicable in the mass range 50~keV to 1~GeV and we find that $f_a$ must be greater than at least $10^{11}$~GeV across this range, peaking at the limit of $f_a\gtrsim 10^{15}$~GeV for $m_a\simeq100$~MeV.  Thus, we have obtained an improved bound over a somewhat smaller mass range.
	
Our bounds provide strong implications for the models which predict heavy-axion-like particles, for example, the axion models in extra dimensions. 
	A detailed analysis of these models is beyond the purpose of this paper, and will be the topic of future work.

\acknowledgments We thank Brenda Dingus for guidance on Fermi observations.

\section*{Appendix: Inclusion of the lepton decay channel}
\label{sec:CommentsOnTheInclusionOfLeptonDecayChannel}

In our calculations we have not included the possibility of HALP decay into leptons, even when this would be kinematically possible ($m_a>2m_\ell$). 	
%%Even if one assumes that the pseudoscalar-lepton coupling is absent in the Lagrangian (e.g. following hadronic axion models), 	
%%the coupling with photons would necessarily induce a (suppressed) coupling with electrons, and so massive axions 	
%%could have another decay channel.  
%This effect would reduce the expected photon flux and somewhat relax our bounds. 
%This result would however be small. 
Here we want to briefly comment on the impact of the axion-lepton decay on our results.

%show that the effect of decay into leptons would however be small.
The effect of a new decay channel would be dual: 
\textit{first}, the axion lifetime would be reduced, and this would affect the exponential factor in 
Eq. (\ref{g_spectrum}); \textit{second}, the decay products would not just be photons. 
This last corresponds to a reduction of the expected photon flux by the factor $\tau/\tau_\gamma$,
where $\tau_\gamma$ is the HALP lifetime into photons given by Eq. (\ref{tau}), and 
$\tau=(\tau_\gamma^{-1}+\tau_\ell^{-1})^{-1}$ is the lifetime including the lepton decay channel. 
For a DFSZ axion,
\[
	\tau_\ell=\frac{6.7\times 10^{10}\,\f^2}{(m_a/\MeV) \sqrt{1 - (2 m_\ell/m_a)^2}}\,\s\,. \nonumber
\]

Both effects would be negligible in the case of a very massive particle.
On the other hand, the reduction factor $\tau/\tau_\gamma$ is small for small axion mass, and could potentially cause a sizable reduction of the photons flux coming from relativistic pseudoscalars. 
However, in general, whenever the condition $t_a/\gamma\tau\ll 1$ is satisfied, the two effects described above cancel out, as can be easily understood by expanding the exponential factor in Eq. (\ref{g_spectrum}). This happens for large values of the PQ constant, and for relatively small axion mass, as confirmed by our numerical analysis in Fig. \ref{fig:leptons}. Therefore, we expect the reduction induced by the lepton decay to be sizable only in the region of intermediate axion mass, and small PQ constant.

	\FIGURE{
	\epsfig{file=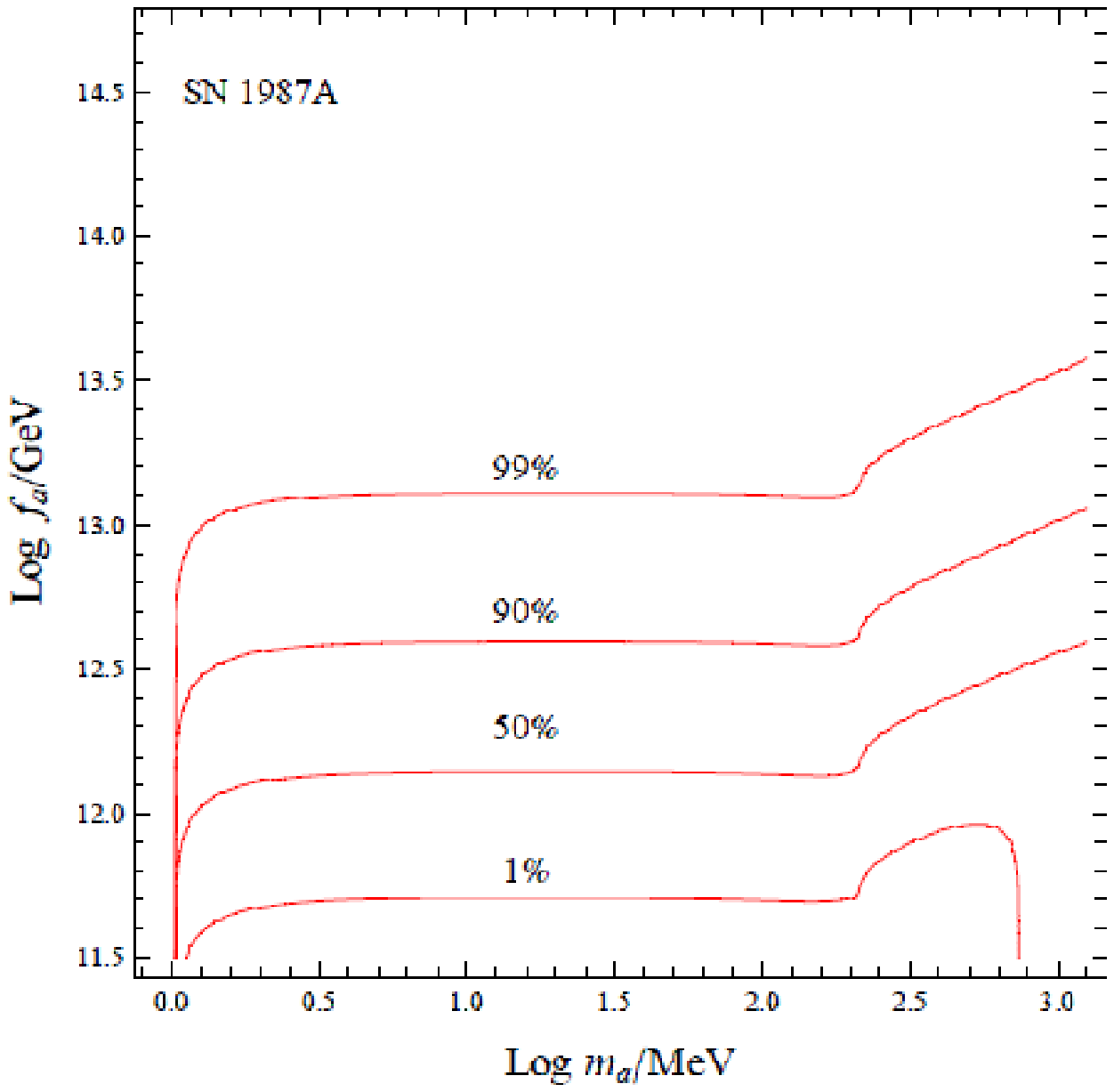,width=0.4\textwidth}
	\epsfig{file=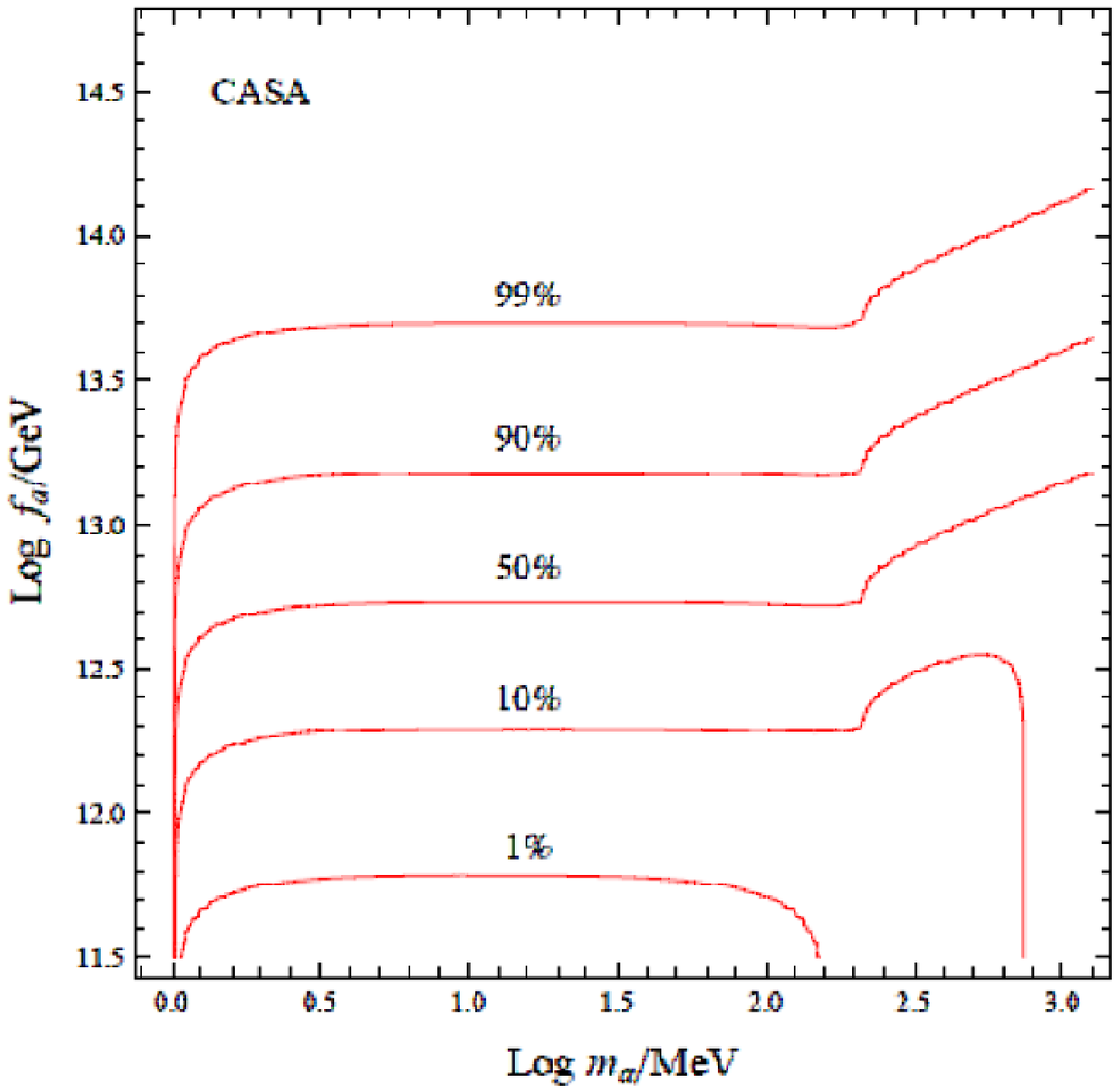,width=0.4\textwidth}
	\caption{Percentage reduction of the flux of photons with energy above 100 MeV expected on Earth 
if the lepton decay channel is open. %In the numerical analysis we are including the decay channel in electrons and muons, and assuming an axion-lepton coupling as for the DFSZ axion. 
	}
	\label{fig:leptons}
	}

Our numerical results are shown in Fig. \ref{fig:leptons}. 
The figure represents the percentage of the flux of photons with energy above 100 MeV expected on Earth 
if the lepton decay channel is open, with respect to the flux (\ref{Eq_F100}), expected if the decay into leptons is neglected.

In the analysis we are including the decay channel into electrons and muons, and assuming an axion-lepton coupling as for the DFSZ axion.
In general, it is possible to consider models where the axion-lepton coupling is much smaller (hadronic axion models), reducing  even more the effects of the lepton decay channel.

Notice that, since $t_a\simeq\Delta t$, the reduction of the flux (at large $\fa$) induced by the lepton decay channel is more pronounced for Cas A, which has a larger delay time.

%However, in the relativistic limit, and in general whenever the condition $t_a/\gamma\tau\ll 1$ is satisfied, the two effects described above cancel out. 

%Analogously, for large $\fa$ the reduction factor $\tau/\tau_\gamma$  
%is compensated by the exponential factor $1-\exp{-t_a/\gamma\tau}\simeq t_a/\gamma\tau$, 
%and negligible reduction is expected.

%relativistic axions, and $\sim 1$ for heavy axion, and does not depend on $\fa$. 
%
%On the other hand, for very large axion mass, the exponential factor in Eq. (\ref{g_spectrum}) is negligible and the multiplicative factor $\tau/\tau_\gamma\simeq 1$. 

%Since we are interest in large values of the PQ constant, the effects of leptons can e neglected.

\bibliographystyle{JHEP}
\bibliography{heavyaxions}

\providecommand{\href}[2]{#2}\begingroup\raggedright\begin{thebibliography}{10}

\bibitem{Peccei:1977hh}
R.~D. Peccei and H.~R. Quinn, {\it C{P} {C}onservation in the {P}resence of
  {I}nstantons},  {\em Phys. Rev. Lett.} {\bf 38} (1977) 1440--1443.

\bibitem{Peccei:1977ur}
R.~D. Peccei and H.~R. Quinn, {\it Constraints {I}mposed by {CP} {C}onservation
  in the {P}resence of {I}nstantons},  {\em Phys. Rev.} {\bf D16} (1977)
  1791--1797.

\bibitem{Weinberg:1977ma}
S.~Weinberg, {\it A {N}ew {L}ight {B}oson?},  {\em Phys. Rev. Lett.} {\bf 40}
  (1978) 223--226.

\bibitem{Wilczek:1977pj}
F.~Wilczek, {\it Problem of {S}trong {P} and {T} {I}nvariance in the {P}resence
  of {I}nstantons},  {\em Phys. Rev. Lett.} {\bf 40} (1978) 279--282.

\bibitem{Kim:1979if}
J.~E. Kim, {\it {Weak Interaction Singlet and Strong CP Invariance}},  {\em
  Phys. Rev. Lett.} {\bf 43} (1979) 103.

\bibitem{Shifman:1979if}
M.~A. Shifman, A.~I. Vainshtein, and V.~I. Zakharov, {\it {Can Confinement
  Ensure Natural CP Invariance of Strong Interactions?}},  {\em Nucl. Phys.}
  {\bf B166} (1980) 493.

\bibitem{Zhitnitsky:1980tq}
A.~R. Zhitnitsky, {\it On {P}ossible {S}uppression of the {A}xion {H}adron
  {I}nteractions. (in {R}ussian)},  {\em Sov. J. Nucl. Phys.} {\bf 31} (1980)
  260.

\bibitem{Dine:1981rt}
M.~Dine, W.~Fischler, and M.~Srednicki, {\it A {S}imple {S}olution to the
  {S}trong {CP} {P}roblem with a {H}armless {A}xion},  {\em Phys. Lett.} {\bf
  B104} (1981) 199.

\bibitem{Kim:2008hd}
J.~E. Kim and G.~Carosi, {\it {Axions and the Strong CP Problem}},
  \href{http://xxx.lanl.gov/abs/0807.3125}{{\tt arXiv:0807.3125}}.

\bibitem{Jaeckel:2010ni}
J.~Jaeckel and A.~Ringwald, {\it {The Low-Energy Frontier of Particle
  Physics}},  \href{http://xxx.lanl.gov/abs/1002.0329}{{\tt arXiv:1002.0329}}.

\bibitem{Raffelt:1996wa}
G.~G. Raffelt, {\it {Stars as laboratories for fundamental physics: The
  astrophysics of neutrinos, axions, and other weakly interacting particles}},
  . Chicago, USA: Univ. Pr. (1996) 664 p.

\bibitem{Duffy:2009ig}
L.~D. Duffy and K.~van Bibber, {\it {Axions as Dark Matter Particles}},  {\em
  New J. Phys.} {\bf 11} (2009) 105008,
  [\href{http://xxx.lanl.gov/abs/0904.3346}{{\tt arXiv:0904.3346}}].

\bibitem{ArkaniHamed:1998rs}
N.~Arkani-Hamed, S.~Dimopoulos, and G.~R. Dvali, {\it {The hierarchy problem
  and new dimensions at a millimeter}},  {\em Phys. Lett.} {\bf B429} (1998)
  263--272, [\href{http://xxx.lanl.gov/abs/hep-ph/9803315}{{\tt
  hep-ph/9803315}}].

\bibitem{ArkaniHamed:1998nn}
N.~Arkani-Hamed, S.~Dimopoulos, and G.~R. Dvali, {\it {Phenomenology,
  astrophysics and cosmology of theories with sub-millimeter dimensions and TeV
  scale quantum gravity}},  {\em Phys. Rev.} {\bf D59} (1999) 086004,
  [\href{http://xxx.lanl.gov/abs/hep-ph/9807344}{{\tt hep-ph/9807344}}].

\bibitem{Dienes:1999gw}
K.~R. Dienes, E.~Dudas, and T.~Gherghetta, {\it {Invisible axions and
  large-radius compactifications}},  {\em Phys. Rev.} {\bf D62} (2000) 105023,
  [\href{http://xxx.lanl.gov/abs/hep-ph/9912455}{{\tt hep-ph/9912455}}].

\bibitem{DiLella:2000dn}
L.~Di~Lella, A.~Pilaftsis, G.~Raffelt, and K.~Zioutas, {\it {Search for solar
  Kaluza-Klein axions in theories of low- scale quantum gravity}},  {\em Phys.
  Rev.} {\bf D62} (2000) 125011,
  [\href{http://xxx.lanl.gov/abs/hep-ph/0006327}{{\tt hep-ph/0006327}}].

\bibitem{Berezhiani:1999qh}
Z.~Berezhiani and A.~Drago, {\it {Gamma ray bursts via emission of axion-like
  particles}},  {\em Phys. Lett.} {\bf B473} (2000) 281--290,
  [\href{http://xxx.lanl.gov/abs/hep-ph/9911333}{{\tt hep-ph/9911333}}].

\bibitem{Rubakov:1997vp}
V.~A. Rubakov, {\it {Grand unification and heavy axion}},  {\em JETP Lett.}
  {\bf 65} (1997) 621--624, [\href{http://xxx.lanl.gov/abs/hep-ph/9703409}{{\tt
  hep-ph/9703409}}].

\bibitem{Berezhiani:2000gh}
Z.~Berezhiani, L.~Gianfagna, and M.~Giannotti, {\it {Strong CP problem and
  mirror world: The Weinberg-Wilczek axion revisited}},  {\em Phys. Lett.} {\bf
  B500} (2001) 286--296, [\href{http://xxx.lanl.gov/abs/hep-ph/0009290}{{\tt
  hep-ph/0009290}}].

\bibitem{Gianfagna:2004je}
L.~Gianfagna, M.~Giannotti, and F.~Nesti, {\it {Mirror world, supersymmetric
  axion and gamma ray bursts}},  {\em JHEP} {\bf 10} (2004) 044,
  [\href{http://xxx.lanl.gov/abs/hep-ph/0409185}{{\tt hep-ph/0409185}}].

\bibitem{Masso:1997ru}
E.~Masso and R.~Toldra, {\it {New constraints on a light spinless particle
  coupled to photons}},  {\em Phys. Rev.} {\bf D55} (1997) 7967--7969,
  [\href{http://xxx.lanl.gov/abs/hep-ph/9702275}{{\tt hep-ph/9702275}}].

\bibitem{Giannotti:2005tn}
M.~Giannotti and F.~Nesti, {\it {Nucleon nucleon bremsstrahlung emission of
  massive axion}},  {\em Phys. Rev.} {\bf D72} (2005) 063005,
  [\href{http://xxx.lanl.gov/abs/hep-ph/0505090}{{\tt hep-ph/0505090}}].

\bibitem{Atwood:2009ez}
{\bf LAT} Collaboration, W.~B. Atwood {\em et.~al.}, {\it {The Large Area
  Telescope on the Fermi Gamma-ray Space Telescope Mission}},  {\em Astrophys.
  J.} {\bf 697} (2009) 1071--1102,
  [\href{http://xxx.lanl.gov/abs/0902.1089}{{\tt arXiv:0902.1089}}].

\bibitem{Pavlov:1999et}
G.~G. Pavlov, V.~E. Zavlin, B.~Aschenbach, J.~Truemper, and D.~Sanwal, {\it
  {The Compact Central Object in Cas A: A Neutron Star with Hot Polar Caps or a
  Black Hole?}},  \href{http://xxx.lanl.gov/abs/astro-ph/9912024}{{\tt
  astro-ph/9912024}}.

\bibitem{Hannestad:2001xi}
S.~Hannestad and G.~G. Raffelt, {\it {Stringent neutron-star limits on large
  extra dimensions}},  {\em Phys. Rev. Lett.} {\bf 88} (2002) 071301,
  [\href{http://xxx.lanl.gov/abs/hep-ph/0110067}{{\tt hep-ph/0110067}}].

\bibitem{Funk:2010cg}
S.~Funk, Y.~Uchiyama, and f.~t. F. L.~C. Collaboration, {\it {Fermi-LAT
  discovery of GeV gamma-ray emission from the young supernova remnant
  Cassiopeia A}},  \href{http://xxx.lanl.gov/abs/1001.1419}{{\tt
  arXiv:1001.1419}}.

\bibitem{Landau}
L.~Landau and E.~Lifshitz, {\it {The Classical Theory of Fields, Fourth
  Edition: Volume 2 (Course of Theoretical Physics Series)}}, . Oxford, UK:
  Butterworth-Heinemann (1980) 402 p.

\bibitem{Sreekumar:1997un}
{\bf EGRET} Collaboration, P.~Sreekumar {\em et.~al.}, {\it {EGRET observations
  of the extragalactic gamma ray emission}},  {\em Astrophys. J.} {\bf 494}
  (1998) 523--534, [\href{http://xxx.lanl.gov/abs/astro-ph/9709257}{{\tt
  astro-ph/9709257}}].

\bibitem{Kolb:1988pe}
E.~W. Kolb and M.~S. Turner, {\it {Limits to the Radiative Decays of Neutrinos
  and Axions from Gamma-Ray Observations of SN 1987a}},  {\em Phys. Rev. Lett.}
  {\bf 62} (1989) 509.

\bibitem{Hanhart:2000ae}
C.~Hanhart, D.~R. Phillips, and S.~Reddy, {\it {Neutrino and axion emissivities
  of neutron stars from nucleon nucleon scattering data}},  {\em Phys. Lett.}
  {\bf B499} (2001) 9--15,
  [\href{http://xxx.lanl.gov/abs/astro-ph/0003445}{{\tt astro-ph/0003445}}].

\bibitem{Kaplan:1985dv}
D.~B. Kaplan, {\it {Opening the Axion Window}},  {\em Nucl. Phys.} {\bf B260}
  (1985) 215.

\bibitem{Turner:1986tb}
M.~S. Turner, {\it {Thermal Production of Not SO Invisible Axions in the Early
  Universe}},  {\em Phys. Rev. Lett.} {\bf 59} (1987) 2489.

\bibitem{Masso:2002np}
E.~Masso, F.~Rota, and G.~Zsembinszki, {\it {On axion thermalization in the
  early universe}},  {\em Phys. Rev.} {\bf D66} (2002) 023004,
  [\href{http://xxx.lanl.gov/abs/hep-ph/0203221}{{\tt hep-ph/0203221}}].

\end{thebibliography}\endgroup

\end{document}